\documentclass[reprint,amsmath,amssymb,aps,prstper,nofootinbib,superscriptaddress]{revtex4-2}
\usepackage{graphicx}
\usepackage{dcolumn}
\usepackage{bm}
\usepackage{hyperref}
\usepackage{xcolor}
\usepackage{comment}
\usepackage{rotating}
\usepackage{colortbl}
\usepackage{enumitem}

\usepackage{graphicx}
\usepackage{dcolumn}
\usepackage{bm}

\begin{document}

\title{Impact of Outreach on Physics Student Development: Qualitative Results from a National Survey}

\author{Jonathan D. Perry}
  \affiliation{Department of Physics, University of Texas at Austin, Austin, Texas 78712}
  \author{Carlee Garrett}
  \affiliation{Department of Physics \& Astronomy, Texas A\&M University, College Station, Texas, 77843}
  \author{Isabella Oaks}
  \affiliation{Department of Physics \& Astronomy, Texas A\&M University, College Station, Texas, 77843}
  \author{James Hirons}
  \affiliation{Department of Physics \& Astronomy, Texas A\&M University, College Station, Texas, 77843}
  \author{Toni Sauncy}
  \affiliation{Department of Physics, Texas Lutheran University, Seguin, Texas, 78155}
  \author{Jonan P. Donaldson}
  \affiliation{School of Education, University of Alabama at Birmingham, Birmingham, Alabama 35294}
  \author{Susan White}
  \affiliation{American Institute of Physics, College Park, Maryland, 20740}
  \author{Rachel L Ivie}
  \affiliation{American Association of Physics Teachers, College Park, Maryland, 20740}
  \author{Tatiana Erukhimova}
    \email{etanya@tamu.edu}
  \affiliation{Department of Physics \& Astronomy, Texas A\&M University, College Station, Texas, 77843}

\date{\today}

\begin{abstract}
The role of student experiences in physics beyond the classroom which support their development has been the subject of exciting research in recent years. Results, typically from small studies at single institutions, have illustrated that facilitating informal physics experiences for non-scientists can enhance student disciplinary identity, learning, sense of belonging, and more. However, it is essential to examine whether these impacts are the sole provenance of institutions with well-developed outreach programs or if they may be shared by institutions anywhere. This work reports on the analysis and findings of responses to three open-ended questions presented to students who indicated they had engaged in facilitating outreach programs as part of a national survey distributed through the Society of Physics Students network in spring 2023. Employing a network analysis with Girvan-Newman clusters revealed six core themes of student experiences: \emph{community participation}, \emph{resilience}, \emph{transformation}, \emph{audience dialog}, \emph{disciplinary development}, and \emph{disciplinary connectedness}. The first four of these clusters were observed to be highly interconnected, providing evidence that the impacts and experiences within them are interrelated with other clusters, particularly interactions with the audience, which is a central feature of informal physics programs. In particular, student experiences highlighted that facilitating informal physics programs enhanced their resilience and belonging, grew their physics identity, provided opportunities to develop essential career skills, and cultivated a growth mindset.

\end{abstract}

\maketitle


\section{Introduction}
Physics departments across the country aim to enrich the educational experience of physics majors and prepare them for their careers. While high-quality coursework and lab activities are central priorities, students often benefit from impactful experiences beyond the classroom. One such experience is engaging students with extracurricular activities such as physics outreach, also called public engagement or informal physics programs, that university students facilitate. These programs can differ significantly in size and frequency \cite{bell2009learning, national2015identifying}, but they give physics majors opportunities to share their enthusiasm for physics and astronomy with the public, helping them see themselves as active members of the physics community \cite{Bergerson2014, hansen2025mobile, clark2016science, Fracchiolla2020, garner2018emergence, garrett2023broadening, Hinko2012, perry2021comparing, randolph2022female, rethman2020creating, tillinghast2020stem, perry2024exploring}.

Less structured than coursework, outreach programs offer physics majors opportunities beyond the traditional focus of physics coursework, such as leadership, creativity, teamwork, and communication. According to studies by the American Physical Society (APS) and the American Association of Physics Teachers (AAPT), these skills have been identified as a high priority for physics programs to prepare students for careers in the 21st century \cite{Heron2016, mcneil2017preparing, mulvey2020physics, pold2015physics, sarkar2016graduate}. In fact, communication remains regarded as one of the main skills that contributes to the employability of students in STEM \cite{Heron2016,leshner2007outreach, noauthor_researchers_2017, foster2010}. A recent survey from the American Institute of Physics's (AIP) Statistical Research group shows that more than 80\% of physics bachelor’s degree holders who graduated in 2019 or 2020 report that they work as a member of a team daily in their job \cite{aip_stats}.


By bringing physics beyond the classroom, students are expected to engage with material in different ways and gain opportunities to interact with a variety of personnel from their home departments. One of the major facets of these programs is students engaging in self-explanation of physics concepts \cite{rethman2020creating, finkelstein2008acting, Hinko2012, Hinko2014, hansen2025mobile} which has been noted to enhance their learning \cite{garrett2023broadening}. Students also gain the opportunity to work side by side with graduate students, researchers, and professors to create and facilitate public engagement which may help them build professional networks and identity. Serving as ambassadors for physics helps them stay motivated and interested in their major. They get recognition as STEM professionals by the general public, their professors, and their peers \cite{rethman2020creating, randolph2022female, perry2021comparing, perry2024exploring, perry2025impactoutreachphysicsstudent}.    

These often impactful and transformative experiences in communicating physics to diverse populations allow students to gain new perspectives and develop self-recognition as physics professionals. By explaining physics to others, students can improve their conceptual understanding. In other words, facilitating outreach physics programs helps students build a disciplinary identity and a sense of belonging within their major and the physics community—factors found to be impactful for student retention and success in the field \cite{Lewis2016, yeager2016using, james2020time, williams2021belonging, dobbins2020time, walton2007question}.

In our pilot study at Texas A\&M University, we surveyed and interviewed students who facilitated five physics outreach programs, ranging from short student-created demonstration videos to a large-scale Physics Festival attended by thousands \cite{rethman2020creating}. Our dataset included 117 survey responses and 35 semi-structured interviews, representing both undergraduate and graduate students. The results were inspiring: students who participated in physics outreach reported growth in their physics identity, a stronger sense of belonging within the physics community, and the development of essential 21st-century career skills. Specifically, they reported improvements in communication, teamwork, networking, and design skills. 

However, our results, as well as other available research, were limited to individual departments and a small number of engaged physics majors. Addressing this gap by surveying a larger number of students engaged in outreach from diverse type of institutions is important. If the Physics Education Research community has compelling data showing that engaging physics majors in facilitating outreach programs can help them build their disciplinary identity and sense of belonging within the physics community, while also enhancing their career readiness by fostering teamwork, communication, and design skills—this could be of significant interest to many physics departments. These benefits, which enhance students' university experience by involving them in outreach facilitation, can be achieved by physics departments of any size without requiring significant financial investment or major curriculum changes. In this paper, we present the results of our first nationwide study. 

Building on our pilot study survey \cite{rethman2020creating} and leveraging collaborative expertise from Texas Lutheran University, Texas A\&M University, the University of Texas at Austin, and the Statistical Research Group of the American Institute of Physics, we developed a new survey instrument incorporating both closed- and open-ended questions \cite{perry2025impactoutreachphysicsstudent}. To maximize responses from undergraduate physics students, we distributed the survey to individual physics majors through the national network of the Society of Physics Students (SPS). This effort resulted in a robust dataset of 704 responses, allowing for a more comprehensive analysis of how outreach programs support undergraduate students than previously available.

The survey aimed to evaluate students' perceptions of their physics identity, sense of belonging, mindset, and career skill readiness, while also collecting data on their engagement with physics outreach programs and demographic background. In our previous paper, we detailed the survey's design, distribution, and analysis, focusing on the closed-ended questions. In this paper, we focus on the analysis of open-ended responses. Using our largest-to-date data base, we investigated the impacts of facilitation of informal physics programs on physics students identity, sense of belonging, mindset, and career skills across a national sample. 

\section{Methods}

To study the impact of student facilitation of informal physics outreach programs, we created a new survey. Drawing on our prior studies at a single institution \cite{rethman2020creating, randolph2022female, perry2021comparing, garrett2023broadening}, the survey combined closed, Likert scale questions and a small number of open-ended questions intended to sample respondents perceptions of their physics identity, sense of belonging, self-efficacy, and career skill development. A more full description of the survey may be found in Perry et al. \cite{perry2025impactoutreachphysicsstudent}. In short, the survey was designed, validated through a pilot distribution to three collaborating institutions, and reviewed for quality, resulting in one minor edit. The survey was then distributed to approximately 5,500 students through the national Society of Physics Students network. 

Respondents who indicated they had participated in informal physics outreach programs received the following prompts. Each such respondent received all three questions and could answer all or none of them. The prompts were:

\begin{itemize}
    \item What has been your most memorable experience from participating in physics/astronomy outreach?
    \item How has participating in outreach provided you with opportunities to utilize your physics/astronomy knowledge in a real-world setting?
    \item How has your perception of yourself as a physicist changed through your participation in physics/astronomy outreach?
\end{itemize}

Analysis of responses to the open-ended questions employed a deductive coding scheme based on theories which had been present during our prior studies \cite{rethman2020creating, randolph2022female}. These theories included a framework for physics identity, incorporating interest and motivation, performance and competence beliefs, recognition, and sense of belonging from Hazari et al. \cite{hazari2020context}. Complementing this physics model, we also included the Dynamic Systems Model of Role Identity (DSMRI) which characterizes context-specific self-perceptions, beliefs, and values \cite{kaplan2017complex}. Additionally, we drew on aspects of situated learning theory \cite{lave1991situated} which described learning through legitimate peripheral participation, transformative learning theory \cite{Mezirow2009, kegan2018form} describing changes to one's frames of reference, and a list of skills identified as essential for 21$^{st}$ century physics careers \cite{Heron2016}. In total, 49 codes were used in this work, organized into the following seven categories: \textit{Identity}, \textit{Community}, \textit{Affect and Experience}, \textit{Disciplinary Skills}, \textit{Non-disciplinary Skills}, \textit{Mindset}, and \textit{Outcomes}.

The category of \textit{Identity} drew from both Hazari et al.'s model and DSMRI and included codes for interest/motivation, internal and external recognition, performance and competence beliefs, curiosity, worldview, legitimate peripheral participation (newcomers learning and integrating into a community of practice \cite{lave1991situated}), and confidence. The category of \textit{Community} incorporated codes of sense of belonging (also from Hazari et al.'s work), connections with peers and audience, accountability at four levels (to the general public, to the scientific community, to outreach leadership, or role based), and impacts from authentic interactions with audience members. The category of \textit{Affect and Experience} included codes for seeing new perspectives, transformational experience, changing assumptions, excitement in facilitators or audience, being uplifted and empowered, desire to persist in the field or major, feelings of authentic purpose or impact from engaging in outreach programs. The category of \textit{Disciplinary Skills} had codes for technical physics or mathematical skills, conceptual understanding of physics principles, developing a sense of real world connections, and engaging in design. The category of \textit{Non-disciplinary Skills} included creativity and innovation, teamwork, leadership, networking, communication with peers or audience, as well as teaching-based skills of scaffolding and zone of proximal development (guiding others in physics beyond their independent abilities \cite{vygotsky1978mind}). The category of \textit{Mindset} was divided into two parts: growth and fixed \cite{dweck2014mindsets}. Each mindset category was further subdivided into three categories of outreach to discipline, discipline to outreach, or ambiguous. These codes indicate whether students attributed their fixed or growth mindset to their experiences in outreach and carried them back to their disciplinary coursework, whether the reverse happened, or if the directionality was unable to be determined. The final category of \textit{Outcomes} incorporated codes based on respondents trajectory towards an academic, a non-academic, or a teaching path. Only responses which were clearly related to experiences from facilitating or working with informal physics outreach programs were coded. Responses for other activities (e.g. summer research or coursework) were not coded.  

Coding of responses was conducted by five researchers. To ensure consistent coding, surveys were divided into subsets of 10-20 which were coded by at least three of the researchers who then met to review codes applied, resolve differences, and refine definitions used. Then all responses were coded. At the end of this process, interrater reliability was $\kappa \geq 0.90$. Work from all researchers was then combined into a single file for analysis. 

The goal of our coding process was to explore the interconnections between the codes and ideas listed above. To achieve this, we employed a network analysis with clustering. This process has been noted to be a valuable tool for investigating the complex relationships between the ideas which comprise the fundamental units (e.g. people, ideas, codes) of qualitative datasets \cite{segev2022semantic, DonaldsonAllenHandy2019}. Starting with the set of coded surveys, a co-occurrence matrix was produced including Pearson's correlation coefficients calculated using MaxQDA Analytics Pro. The co-occurrence matrix examines pairs of codes for both where they do and do not appear together in the text of documents within the dataset. To limit this work to the most significant relationships between codes, we examined correlations only with a significance level of p$<$0.001. The matrix of correlational data is then used as input to create a 1-mode network using the UCINET software \cite{UCINET2002}. This software generates a visual representation of significant relationships (edges) between nodes (codes). Size of nodes within this map are determined using eigenvector centrality, which accounts for both the strength and number of significant correlations, yielding larger nodes for codes which are more central to the network \cite{Kadushin2012}. To further narrow the number of codes to a small set of core themes, we also employed a Girvan-Newman clustering algorithm \cite{girvan2002community}. This process examines the betweenness of codes, subdividing a potentially complex map into a reduced number of regions where codes are more closely related to each other than they are related to the rest of the map. Clusters may then be considered to represent macro-codes, or over arching ideas within the network map. The clustering algorithm output is considered robust for values, $Q$, above 0.30 \cite{zahiri2023improved}.

\begin{table}
    \caption{\label{tab_demo} Demographics of students who responded to at least one open-ended question in the national survey. In some cases, percentages do not add to 100\% as respondents could select `I prefer not to respond' or a similar non-identifying answer. For ethnicity, respondents were allowed to choose multiple answers and categories have been condensed to two to preserve anonymity during the analysis phase.}
    \begin{ruledtabular}
        \begin{tabular}{ll}
            Demographic & Percent \\ \hline
            \textit{Gender} & \\
            Men & 49\% \\
            Women & 41\% \\
            Another Identity & 7\% \\ \\
            \textit{Institution Type} & \\
            Bachelor's & 37\% \\
            Master's & 4\% \\
            PhD & 46\% \\ \\
            \textit{Race or Ethnicity} & \\
            White & 62\% \\
            Person of Color & 27\% \\ \\
            \textit{LGBTQIA+} & \\
            Yes & 25\% \\
            No & 63\% \\ \\
            \textit{International Student} & \\
            Yes & 8\% \\
            No & 90\% \\ \\
            \textit{Physical or Mental Disability} & \\
            Yes & 47\% \\
            No & 48\% \\
        \end{tabular}
    \end{ruledtabular}
\end{table}

\section{Results}
\begin{figure*}
    \centering
    \includegraphics[width=0.95\linewidth]{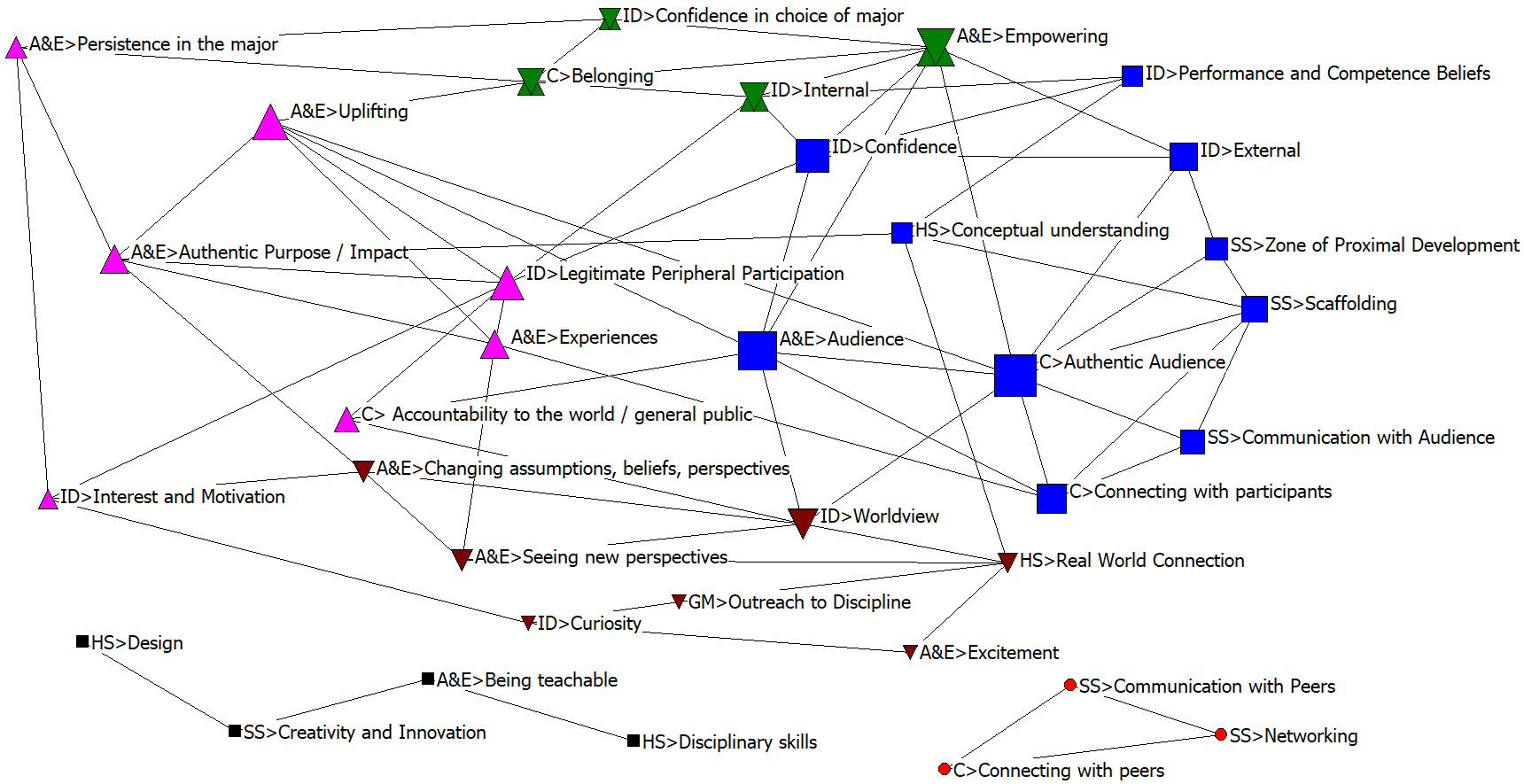}
    \caption{Network map at the p$<$0.001 level of relationships between codes included in this study. Codes with no statistically significant relationships to other codes are omitted. The size of codes is representative of their eigenvector centrality. Codes are grouped into clusters (by color and shape) using a Girvan-Newman clustering analysis with $Q=$ 0.476.}
    \label{network_map}
\end{figure*}

In this section, we describe our findings from a network analysis of coded responses to the three open-ended questions included as part of the survey. Of the 704 respondents to the survey, approximately 70\% participated in some form of outreach. Of those, 239 responded to one or more of the three open-ended questions yielding 621 non-trivial responses. The demographic information of the respondents is shown in Table \ref{tab_demo}. Overall, respondents represented a reasonable balance of demographics across the spectrum of undergraduate physics majors. Results of the network analysis, at the p$<$0.001 level, are shown in Fig. \ref{network_map}. Results from the Girvan-Newman clustering algorithm show six distinct clusters, distinguished by color and shape, with $Q$=0.476. The research team assigned a name to each cluster which represents the core theme of the interrelated ideas. Below, we describe and summarize each of these clusters, with supporting quotes and analysis from the survey responses. All quotes are shared verbatim unless indicated with square brackets to denote minor word changes to fit the sentence structure.
 
\subsection{Clusters}

The cluster represented by the green hourglass icons was labeled \emph{resilience}. Within this cluster appears the codes: belonging, confidence in choice of major, empowering, and internal recognition. Experiences which related to these codes heavily favored internal shifts in students' perceptions of themselves within the departmental community and how certain they were that they had picked the right major or area of study for themselves. The role of outreach was in providing opportunities for students to reflect on their own abilities and learning, as well as their social connectedness and integration within the physics community. 

After participating in outreach, one student in particular summarized these impacts in informal physics programs thusly: ``I feel better involved in my major, and I feel more self-assured as a physicist. I truly feel like I have picked the right major that aligns well with what I can do and what I am interested in doing.'' Another student went into greater detail, sharing how in their freshman year they felt like ``just another physics major'' and that they perceived themselves at that time to be ``nothing special''. However, once they started facilitating outreach this idea of ``not being special or notable was completely wiped out'' of their mind. They ``no longer felt like just another average physics student. Outreach gave [them] a sense of belonging and purpose.'' Their experience led them to be recognized by others in the department, both faculty and students, and become a physicist that the department could depend on. This highlights a significant shift in their internal perceptions of themselves, which they attribute heavily to outreach, culminating in a sense that they are ``worthy of the title of a physics major.'' Other respondents in this cluster were briefer in their comments but repeatedly identified that outreach validated, reinforced, or affirmed their choice in major. In the words of one student ``if I had to do life over again, I'm pretty confidence that would choose physics again.''

The cluster comprised of dark red inverted triangle icons was labeled \emph{transformation}. This cluster is comprised of the codes: curiosity, worldview, excitement, changing assumptions, beliefs, and perspectives; seeing new perspectives, and growth mindset - outreach to discipline. As a whole the codes relate to shifts in students outlook on their path in physics, a redefining of what it means to be a physicists, or about the malleability of their skills within the discipline (indicative of growth mindset). In regards to outreach, respondents attributed these changes to their intrapersonal or interpersonal experiences related to outreach events and programs. 

For some students, their engagement in informal programs acted as a way to put their current knowledge in context. As one student shared ``I realize[d] how little I actually know and wish to expand that knowledge.'' While others began to see themselves in different ways than they had before. This was exemplified by students sharing that they ``still have a ways to go, but [they] have grown to see [themselves] as someone who shared knowledge with others'' or that they were ``reminded how far [they] have come in [their] understanding of physics/astronomy'' when communicating it to others. For other students this change in their outlook addressed their feelings of imposter syndrome. In particular, multiple students shared that by communicating physics to the public it had helped them to ``come to terms'' or ``alleviate'' their imposter syndrome. One respondent in particular elaborated on how they ``did not feel the need to compare [themselves] to the incredibly bright people in [their] department. [They] feel more secure in [their] identity as a physicist.'' Outreach also had an impact on students worldview, helping some to ``understand the vitality of science communication'' and how ``valuable a skill'' that can be for a physicist. Through growing that skill set students may experience personal growth where they ``no longer feel like just a student in the physics community, but also an active member of it.''

The cluster of blue squares was termed \emph{audience dialog}. This cluster is comprised of codes: authentic audience, audience excitement, confidence, external recognition, performance and competence beliefs, zone of proximal development, communication with audience, connecting with participants, scaffolding, and conceptual understanding. Centered on engagement with an authentic and curious public, this cluster represents both changes to student self-perceptions as physicists stimulated through this engagement as well as the development of essential career skills around communication. Student shifts in self-perception were focused on awareness of learning physics concepts, and on aspects of physics identity, specifically external recognition.

Students often reported on their sense of excitement and joy at interacting with the public. From feeling ``inspired'' by kids asking ``fantastic questions'', or seeing ``fascinated little faces'' when engaging with physics demonstrations, to touching exchanges where a middle school students went up to a facilitator and shared that ``he wanted to be a scientist when he grew up'' after having so much fun; respondents demonstrated how nearly ubiquitous positive interactions were during informal physics outreach programs. Pushing deeper than the excitement, respondents went on to share their awareness of how they had changed by working with their outreach programs. Regarding career skill development, one student shared how outreach ``helped [them] find the correct vocabulary that the general public is comfortable with.'' Another contextualized their outreach experience as helping them learn the best ways to interact ``with those who both know and don't know'' physics. This also helped to push students to deepen their own understanding of the topic since, as one student put it, having the ``opportunity to explain physics in a simplified manner means I need to understand it well to that it makes sense to younger children or people without science background.'' 

These interactions through authentic exchanges were also observed to help students alter their views of themselves due to the situation and feedback received. From one respondent's most memorable experience they recalled ``talking to one child, being completed engrossed in the explanation, and not noticing until [they] had finished [their] explanation that this was the child of one of [their] professors! [They] gave a much better explanation and demonstration because [they] didn't realize that there was potentially someone else watching over [them], leaving [them] more relaxed even when talking to influential people in the field.'' Another student shared that after having showed a young girl several demonstrations she went on to ``reteach [the respondent] everything [they] had taught'' the young girl. Her take away from this experience was that she ``was not only making a difference but proved that [they were] smart and good enough to make a difference.''

The cluster of pink triangles was termed \emph{community participation} focused on respondents experiences specifically with their physics community. The codes within this cluster include: legitimate peripheral participation, authentic purpose/impact, interest and motivation, persistence in the major, accountability to the world/general public, uplifting, and transformational experiences. Respondent experiences within this cluster centered around integration into the overall physics community through their experiences in outreach, as well as transitions of self-perception and persistence in the field. In particular, a theme of self-perceptions centered on shifts due to accountability, students feeling a duty to give back to the world through their positionality as a member of the physics community. It is notable that, though not present within the network map at the level of significance used for this study, this cluster contains a majority of responses where students referenced a focus on teaching as a future profession. 

By engaging in informal physics programs, respondents discussed their perceptions of themselves shifting through their engagement with others. One student in particular highlighted that ``even though [their] classes are difficult, which can be discouraging, [they] are reminded how far [they] have come in [their] understanding of physics/astronomy when [they] are teaching others about physics/astronomy topics. This helps [them] feel more like a physicist and less like a floundering student.'' Others were more succinct simply stating that ``more and more, I see myself as a physicist'' or that outreach helped to alleviate imposter syndrome and helped them to ``feel more secure in [their] identity as a physicist.'' Some respondents also spoke of the connectivity they gain through outreach. From one student's experience their ``participation in physics and astronomy outreach has added depth to [their] identity as a physicists. It keeps [them] connected to the community.'' For at least one respondent, their experiences in outreach provided opportunities to form essential relationships to broadened their roles within the department as outreach ``made [them] feel not only as a student, but also a mentor and a mentee to professors and new students that are starting in the field.'' This is echoed by another response which shared how the respondent realized that they ``may be a role model for some.'' Numerous responses also discussed a feeling of necessity of giving back to the general public as an essential element of being a physicist. As one response framed it, outreach changed their perception ``from wanting to just work on the nitty-gritty of research and science collaboration, to acknowledging the great importance of being able to communicate those complex topics to a general audience.''

An additional theme from this cluster showed that through their community engagement a number of respondents experienced a growth or reinforcement of their desire to teach physics in the future. After working with outreach programs some respondents shared that they could ``actually see [themselves] teaching physics in the future'' or that working with elementary aged students and seeing the looks on the kids faces ``really validated [their] goal of becoming a physics teacher.'' One in particular shared how through outreach they ``found that [they] liked talking about research far more than doing research'' and this had helped them ``chose teaching as [their] ultimate career goal.''


The cluster comprised of black squares was labeled \emph{disciplinary development}. This cluster contains codes: design, creativity and innovation, being teachable, and disciplinary skills (related to applying physics knowledge). Together these codes relate to constructionist impacts through creation activities which contributed to enhancing students physics knowledge. A majority of experiences shared by respondents centered on their work building or creating demonstrations for use in outreach events. 

This cluster is perhaps best summarized by one student's experience that ``Outreach has made me more confident in my perception as a physicist. I am able to create something, learn from its creation, and then present my knowledge to the public.'' Another student adds to this idea, that creating a demonstration required them to see ``how feasible and practical the theoretical knowledge'' can be. Here we see impressions of the direct link between creating and learning, as well as development of confidence seen in another cluster. Through the eyes of other respondents, this process of creation was also seen to ``not only test my physics knowledge but also my creativity'' or that ``turning an idea into a tangible, working creation and the work involved in that allowed me to utilize my physics knowledge in a real-world setting.'' These activities are then seen to further physics knowledge by developing creative new solutions which must of course ``effectively communicate the concepts''. This process is not necessarily a quick one either, potentially overflowing from a traditional semester as seen with one student who ``had spent the year designing'' a significant astronomy-based activity. While not all facilitators of outreach engage in a design process or create new demonstrations, some do. Students reported that this process helped them be cognizant of how they are ``absorbing the knowledge'' from classes and allowing them to push their boundaries ``getting practical knowledge'' in bringing a demonstration to life, deepening their understanding of specific physics topics while developing their skills of creativity to share something with their audience. 

The final cluster identified by red circles was termed \emph{disciplinary connectedness}. This cluster is comprised of three codes all interconnected: networking, communication with peers, and connecting with peers. Combined, these codes related to the experiences of respondents in getting to meet and know their classmates, professors, staff, and other departmental personnel in less formal contexts working through outreach events. 

This cluster seems to have had two somewhat different impacts on students. For some, engaging in outreach helped them to become ``[a] more socially capable person that can talk more easily to other physics majors''. Multiple respondents discussed the development of their social circle through their outreach connections, highlighting that it helped them to make ``so many friends in the department'', to ``feel better involved with [their] major'', or that the ``relationships [they'd] built with other participants'' had been their most memorable experience through outreach. For others, their experiences through outreach had a more direct impact on their learning via enhanced interactions related to coursework. As one student shared ``without the confidence I gained through outreach, I don't think I would have asked as many questions in class, spoke up in group activities, or had as much confidence in the work I produced. Outreach taught me how to effectively communicate my thoughts to professors and fellow classmates, this skill alone contributed to a lot of my success as a physics major.'' By providing an opportunity to develop the social connections within their discipline this student was able to better engage and speak up with others in the department, enhancing their overall learning. Through outreach, students can get opportunities to make authentic connections with their classmates, with students from other years in their majors, with faculty, and potentially with graduate students and post-docs depending on their institution types.

\begin{figure*}
    \centering
    \includegraphics[width=0.95\linewidth]{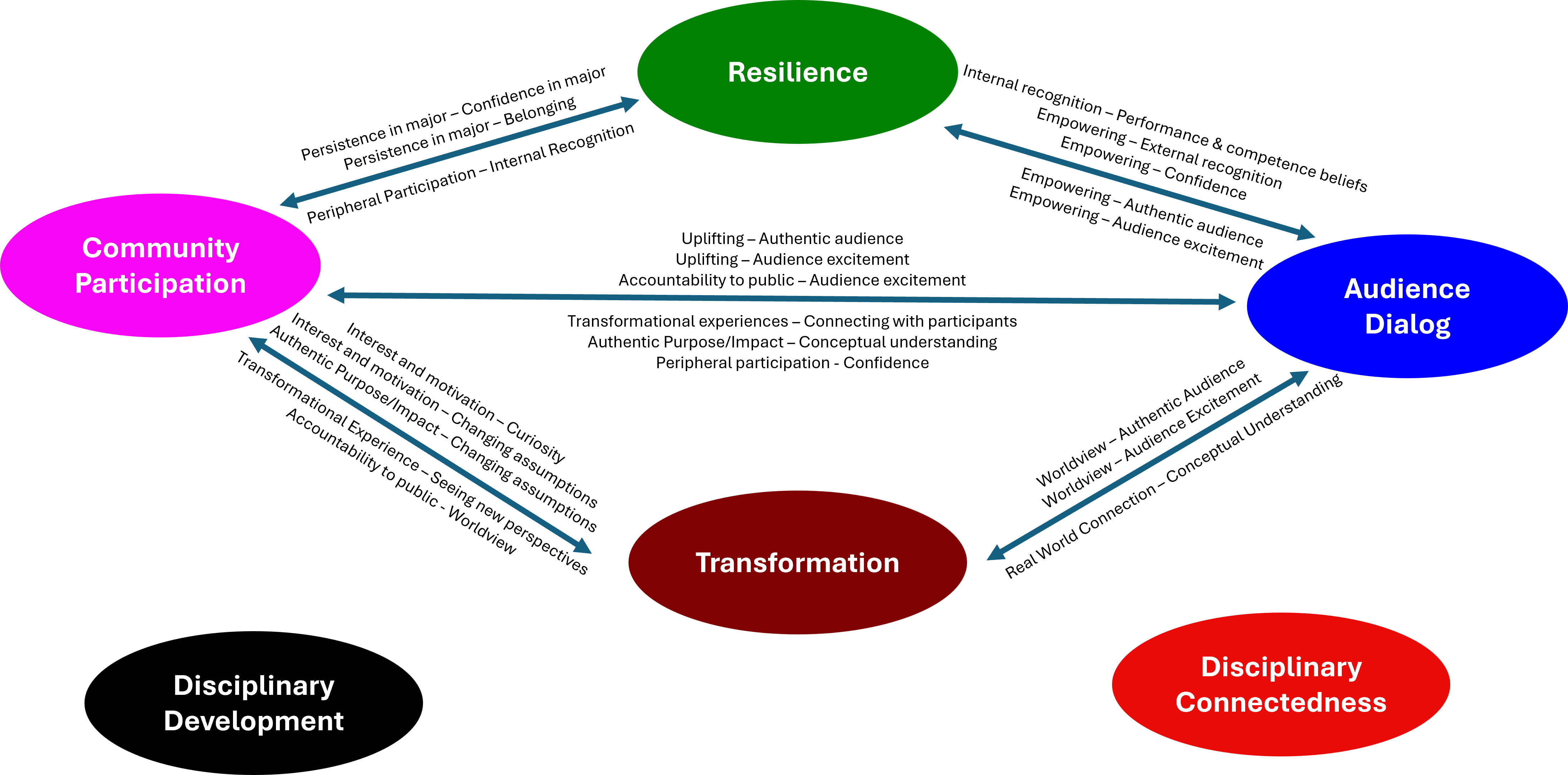}
    \caption{Simplified version of the network map highlighting the central theme of each cluster and the edges through which the interrelated clusters were connected.}
    \label{reduced_network_map}
\end{figure*}

\subsection{Connections}

From Fig. \ref{network_map} it may be seen that four of the six clusters are connected across multiple edges. A simplified network map highlighting these connections is shown in Fig. \ref{reduced_network_map}. These connections demonstrate that the impacts of students engaging in informal physics programs are, mostly, not independent but are interrelated. Here, we briefly highlight the experiences of students at the intersections of these four clusters. It is interesting to note that no statistically significant links are present between the \emph{transformation} and \emph{resilience} clusters. 


The \emph{community participation} and \emph{resilience} clusters connect along three edges: Persistence in the major - confidence in choice of major, Persistence in the major - belonging, and Legitimate peripheral participation - internal recognition. Experiences from students which exemplified these connections focused on the importance of engagement within the physics community reinforcing their sense of belonging in physics and/or a desire to complete their studies in the field. Students reported that through engaging in outreach they felt ``more accepted in the physics community and also [felt] more motivated to keep working hard on [their] dreams and projects.''. Another shared that facilitating outreach helped them feel as if they ``belong[ed] more amongst [their] peers after getting to be around them and it's help[ed] give [them] the confidence and drive to complete the major.'' Through working with their peers students were able to engage with colleagues who shared their interests and values which made them ``feel like [they] belong in physics'' and that ``the more [they] feel like [they] belong the more comfortable [they] feel calling [them]selves a physicist.'' These experiences allowed students to feel ``more involved in [their] major than before'' after facilitating outreach, allowing them to feel comfortable saying ``that [they are] a physicist finally.''


The \emph{resilience} and \emph{audience dialog} clusters connect along five edges: Internal recognition - performance and competence beliefs; Empowering - external recognition; Empowering - confidence; Empowering - authentic audience; and Empowering - audience excitement. Often respondents discussed becoming ``more confident'' in their knowledge and abilities. They felt ``more of a professional in the field when [they] do more outreach programs.'' The opportunity to engage with audiences provided a valuable perspective of student capabilities in physics leading to more certainty about persisting in the field. For some this came about as a reminder that ``physics is a source of genuine interest'' and that the validation provided by the general public reminded them that they are ``not just someone with a passing interest'' in physics. These experiences were particularly valuable to one student who felt they did not fit into their department, stating ``there are very very few people who look like [them] in [their department]'', but that every time ``[they met] someone like [them] who's interested in doing astronomy, [they] feel a little bit more like [they] belong.'' One student summarized the impact of this connection very well, sharing that ``watching other people's eyes light up when [they] explain something is a powerful feeling''. This made them ``feel more confident since [they] can clearly see that [they're] capable of explaining'' physics. They went on to say that going back over concepts ``made [them] feel much more solid in [their] existing knowledge, and has reaffirmed that [they] haven't gotten dumber as much as the material has gotten more complicated.''


The \emph{community participation} and \emph{audience dialog} clusters connect along six edges: Uplifting - authentic audience; Uplifting - audience excitement; Accountability to the general public - audience excitement; Transformational experiences - connecting with participants; Authentic purpose/impact - conceptual understanding; and Legitimate peripheral participation - confidence. These links represent perceptions of the progressive integration of students into the physics community through their interactions with the general public through informal physics programs. Many respondents shared inspiring stories of their interactions with audience members, often young children, helping to make them feel more part of the physics community. This could be as simple as ``helping younger children see how interesting physics and astronomy is'' to specific instances that stuck with respondents. One such example is from a student ``working with a high school girl who didn't even realize you could make a career in physics and astronomy. She is is now pursuing a physics major in her first year of undergrad and [the respondent is] so grateful [they] got to show her that field.'' These interactions can also serve as a measure of student progress within the discipline. One student shared about their experiences in a pen pal program seeking to reduce stigma around physics and astronomy. These exchanges allowed the student to both ``watch how much progress [they're] making towards that goal and [they] get to help a student regain some amount of passion for science'' which they may lose through their regular schooling. 


The \emph{community participation} and \emph{transformation} clusters connect along five edges: Interest and motivation - curiosity; Interest and motivation - changing assumptions; Authentic purpose/impact - changing assumptions; Transformational experiences - seeing new perspectives; and Accountability to general public - worldview. Numerous respondents reported that outreach helped deepen their passion for their chosen subject, or that they ``realized how much [they] love physics.'' It also helped some to refine their ideas about what a physicist should be, not just spending all of their time in a lab or engaging in the scientific process but also helping non-physicists to understand the process and its results. From one student's perspective outreach helped to show the ``disconnect between the way we approach science in academic and the general public's interest in learning fun science facts.'' The student continued saying that it ``has helped [them] to take [themselves] less seriously, as... physics is a source of genuine interest for [them] and isn't always just another academic challenge.'' Others went further describing that outreach developed a sense of accountability or need to ``communicate and imbue knowledge onto the general public'' and that as scientists we should be ``helping the general public to understand results and ensuring that science is for everyone''.


The \emph{transformation} and \emph{audience dialog} clusters connect along three edges: Worldview - authentic audience; Worldview - audience excitement; and Real world connections - conceptual understanding. Interactions with the general public helped to facilitate a transformation of respondents' self-perceptions. This was driven both by direct feedback from members of the audience, as one respondent shared they ``didn't realize how people viewed physicists so highly'', or by bringing physics off of the board and into reality making it ``not just a lecture or homework but something that's real.'' Through these interactions students observed themselves becoming more comfortable being in a position to share their knowledge and finding ways to make it relevant for those outside the discipline. As one student shared outreach ``has shown [them] that there is physics all around us'' and they ``see it everywhere'' they go which leads them to explain the physics to peers even if they aren't familiar with it. After outreach they can ``more easily visualize what physics problems may be describing, and [their] ability to interpret more complex topics has become much easier.'' Another reported a similar shift in their abilities stating that they are ``now so comfortable explaining a physics concept, that [they] can generally ask someone what they do and find somewhere in their life or career that the specified concept is applied to help them understand it better.'' These interactions can also lead to a shift of values for future careers. As for one student, talking to non-scientists ``really pushed [them] to consider how [their] physics knowledge can be useful outside of academic'' leading them to consider how their work might ``improve the life of the average person''.

\section{Discussion}
As a discipline, physics is often known for having high attrition rates for students who enroll in the program. A swath of literature show that important factors which can impact students choices to persist, or not, within the major relate to their physics identity, sense of belonging, and mindset. Recent work in both K-12 and higher education have noted this effect in other STEM fields as well, particularly when student experiences are limited to courses without pedagogical practices designed to support the development of these constructs \cite{hecht2022peer, walton2023and, yeager2022synergistic, burnette2023systematic}. When such practices are implemented, students will often experience improved grades, take more challenging classes, and become more resilient to struggle as they move through their education \cite{yeager2016teaching, yeager2016using, baldwin2020promoting}. 

As follows from our results, engagement in outreach provides venues for students to find their disciplinary voice, make individual contributions to their home departments, and feel that they belong. By drawing students to share their passion for and knowledge of physics beyond the classroom, they are able to engage in authentic activities with the general public which provide their own positive messaging allowing students to develop their disciplinary identity and sense of belonging. It is notable to observe that the four elements of Hazari et al.'s physics identity model (interest and motivation, internal and external recognition, sense of belonging, and performance and competence beliefs)  \cite{hazari2020context}  are embedded across three of the four interconnected clusters (\emph{community participation}, \emph{resilience}, and \emph{audience dialog}) in Fig. 1. These results echo work examining Learning Assistant programs which are also seen to enhance student identity and belonging through their activities as near-peer instructors in, and outside of, the classroom \cite{Close2016}. 

It is worth reiterating on the presence of the \emph{resilience} and \emph{community participation} clusters within the network as this has important implications for attrition, or persistence, in physics. Working through informal physics program provided unique experiences to students which reinforced their confidence in their choice of major and desire to persist in it, mirrors results from both university and high school settings \cite{randolph2022female, Perez2014, Hazari2010}.

Experiences highlighted in the \emph{transformation} cluster demonstrate how outreach allowed students to develop their own ideas of what it means to be a physicist and how they should engage with the world around them. In particular, this came in parallel to students noticing a reduction to their feelings of imposter syndrome, something common to scientists and professionals at all levels \cite{chakraverty2020impostor, beesley2024undergraduate, ivie2016women}. These results mirror our prior work which saw similar impacts for students who had engaged in outreach \cite{randolph2022female, rethman2020creating}. Through broadening their engagement beyond their professors and classmates, students receive essential feedback that clearly demonstrates the progress of their knowledge and understanding which can assault the foundations of imposter syndrome. 

The impacts of engagement with a general audience are not limited to mitigating imposter syndrome. As seen in Fig. 1, the cluster of \emph{audience dialog} is significantly connected with multiple other clusters. Respondent experiences quoted in the Results section regularly attributed the changes in constructs related to their disciplinary identity, belonging, and mindset to the authentic engagement, passion, and recognition they receive from audiences of all ages. Therefore, there is a clear link between the material benefits of student facilitation of informal physics outreach programs with the general public. The impacts observed are an intrinsic feature of these interactions and may not be replicable within a formal curriculum. This importance of audience interactions have been noted by multiple prior studies at single institutions \cite{rethman2020creating, Hinko2016}.

Feedback from audiences had multi-faceted impacts noted in the prior section, being significantly linked to resilience, community participation, and transformation. These interactions allowed for significant opportunities for students to hone and expand their communication skills and to receive strong feedback signals that demonstrate to students how their abilities and knowledge grew with time. By engaging with people beyond the discipline, students received significant opportunities to improve their communication skills, working to make themselves clearly understood at the level of their audience. This vital career skill has been identified as essential to practices for physics students in professional positions \cite{pold2015physics, mulvey2020physics}. These skills are transferrable to any profession, whether in industry, government, or academic environments. Indeed, according to work by Sarkar et al. \cite{sarkar2016graduate} communication skills, among others, were noted as being missing or deficient in training within the major for recent graduate students. Within the context of outreach, students receive authentic opportunities to work on valuable skills including analogies, illustrations, and explanations \cite{Hinko2012} which enhance dialog and understanding with non-technical audiences. This improvement to communication skills has been noted in prior work \cite{rethman2020creating, Hinko2014, randolph2022female} as a result of outreach facilitation. It is important to note that communication was strongly linked to conceptual understanding. As seen in our prior work from Garrett et al. \cite{garrett2023broadening}, by engaging in the explanation process students refine their own individual ideas and improve their depth of physics knowledge. 

Other skills noted as being undertaught in the physics curriculum include those of team-working, organization, and adaptability \cite{sarkar2016graduate}. Figures 1 clearly indicates students perceiving themselves as improving these skills through their outreach activities. By engaging in the process of design and creation, often for developing an apparatus or demonstration for events, students found themselves honing their creativity and ability to innovate. Outreach is a particularly fruitful place to engage in these activities as coursework and even advanced lab work can tend to be convergent, allowing for only one or two unique approaches. Allowing students to stretch themselves beyond canonical work can provide vital opportunities for discovery, and even low stakes failure, that can be valuable for future careers. Further, the social aspect of outreach allows for significant connections to form between students and their classmates, faculty, graduate students, or other physics personnel. These relationships can, among other benefits, provide opportunities for mentorship in the moment, recently noted as a vital aspect of quality education across their time as students \cite{felten2020relationship}. This is similar to what was observed in our prior work where students reported significant opportunities to work on their creativity through building demonstrations in teams and being able to develop their personal networks across the department \cite{rethman2020creating, randolph2022female}.


Appearance of the code of growth mindset on the network map provides us with an important hint. In our quantitative study, we observed a statistically significant link between student engagement in outreach programs and signals of growth mindset. The regression models which demonstrated this link were unable to provide information about the directionality or cause, that is which way the influence went. This left us with an uncertainty as to the flow of the impact, whether students with a growth mindset were more likely to participate in outreach or students who participated in outreach were more likely to develop more of a growth mindset. The growth mindset code in Fig. 1 represents the sub code of directionality of outreach to discipline. It is notable that this is the only one of the six mindset codes from the code book which appears within the network map. We assert that this signals a likely explanation for the direction of the effect from our previous study that students facilitating informal physics programs are more likely to develop a growth mindset. Respondent experiences demonstrate a consistent trend that engaging in informal physics programs helped them to better understand the malleability and evolution of their physics knowledge and abilities. Several responses indicated that outreach helped students develop their growth mindset. They elaborated to share that they felt better able to tackle the challenges of their formal curriculum after this change. This result echoes our previous study which observed women who had engaged in outreach strongly reported signals of growth mindset \cite{randolph2022female}.

\section{Limitations}
The results from open-ended questions posed as part of the national survey exhibit interesting relationships and themes. However, we must appropriately note some limitations in interpreting these findings. All data was self-reported from students responding to the survey. Further, there is potentially a selection bias as part of the optional survey sent to students, where only those from the extremes may have opted to complete it.  

\section{Conclusion}
Drawing on the largest national sample available to date, we examined the impact of student facilitation of informal physics outreach programs on constructs of physics identity, sense of belonging, mindset, and career skill development. Students responding to the survey were drawn from SPS chapters throughout the country incorporating experiences from institutions with variably developed informal physics programs. Results from a mixed-methods analysis of responses to three open-ended questions posed to students who had engaged in outreach activities broadens our understanding of impacts on student physics identity, sense of belonging, mindset, and career skill development with notable parallels to results from our prior single institution study \cite{rethman2020creating}.

Analysis of 239 respondents answers to open-ended questions relating to their experiences about engaging in informal physics outreach programs has revealed an important set of themes centered on six clusters. These clusters were \emph{resilience}, \emph{community participation}, \emph{transformation}, \emph{audience dialog}, \emph{disciplinary development}, and \emph{disciplinary connectedness}. The first four of these clusters exhibited multiple statistically significant links, indicating that the impacts described within are not independent of each other. In particular, student experiences of being able to share their physics knowledge and passion with a general audience were linked to higher confidence in their choice of major, sense of belonging to the physics community, and, as a result, characteristics of resilience. Students also felt more integrated into the physics community and practices by being able to inhabit expert-like positions, and were seen to share higher levels of intrinsic motivation to learn more about their chosen field. Constructs of physics identity \cite{hazari2020context, kaplan2017complex} were distributed through multiple clusters. These codes were heavily interconnected with codes relating to engaging with audience members, indicating that the benefits to enhancing students' physics identity appear to come directly from the nature of outreach which provides flexible ways to engage with non-scientists through low stakes and less structured interactions. 

The work of a physicist after graduation relies heavily on skills beyond that of solving physics problems or lab related work. To be effective in the workplace, physics students in the 21$^{st}$ century are expected to gain a broad range of skills including teamwork, communication, design \cite{Heron2016, sarkar2016graduate}. Results indicate that students gain significant opportunities to broaden and improve their skill sets through facilitation of outreach programs, particularly when it comes to communication to non-technical audience, creativity, and networking. Through opportunities to work alongside colleagues in low stress environments, students form potentially long lasting bonds while they explore ways to polish their methods of sharing abstract and challenging ideas in accessible and engaging ways, which often led to improved conceptual understanding. These are skills which are easily transferrable to whatever endeavors students pursue after graduation. 

Whereas a student just entering the field may see the work of a physicist as one-dimensional efforts in a lab or a chalkboard, those more familiar with the field recognize how limited this view is. Our results indicate that by providing opportunities for transformation, outreach allowed students, early on, to see the breadth and complexity of the work of a physicist and permits them to explore a range of aspects of the field to find what they most identify or resonate with. A particularly important aspect of this is seen by the growth mindset signal which indicates that outreach enriches this perspective of malleability that students can bring back to the classroom.


\section{Acknowledgments}

This work was supported by NSF Grant No. IUSE 2214493. We would like to thank Brad Conrad for his assistance in distributing the survey and John Tyler for his work in preparing the data and analyzing quantitative results.


\bibliography{apssamp}

\end{document}